\documentclass[reprint,
					aps,
					pra,
					english,
					footinbib,
					10pt,
					superscriptaddress
]{revtex4-1}

\usepackage[english]{babel}

\usepackage[utf8]{inputenc}

\usepackage[T1]{fontenc}

\usepackage[intlimits]{amsmath}

\usepackage{amssymb}

\usepackage{mathtools}

\usepackage{dsfont}

\usepackage{siunitx}

\usepackage[final,colorlinks=true,linkcolor=blue,citecolor=blue]{hyperref}

\usepackage{appendix}

\renewcommand*\d[1]{d #1\,}
\newcommand{\ii}{\mathrm{i}}
\DeclarePairedDelimiter{\abs}{\lvert}{\rvert}
\newcommand{\ee}[1]{\operatorname{e}^{#1}}
\newcommand*{\defeq}{\mathrel{\vcenter{\baselineskip0.5ex \lineskiplimit0pt
	\hbox{\scriptsize.}\hbox{\scriptsize.}}}%
=}

\newcommand{\ket}[1]{\vert #1 \rangle}

\DeclarePairedDelimiterX\braket[2]{\langle}{\rangle}{#1\,\delimsize\vert\,\mathopen{}#2}
\DeclarePairedDelimiterX\matrixel[3]{\langle}{\rangle}{#1\,\delimsize\vert\,\mathopen{}#2\,\delimsize\vert\,\mathopen{}#3}

\DeclareMathOperator\perm{perm}

\newcommand{\includefig}[1]{\includegraphics{#1}}

\usepackage{verbatim}

\begin{document}
\title{Scattershot multiboson correlation sampling with random photonic inner-mode multiplexing}

\author{Vincenzo Tamma}
\email{vincenzo.tamma@port.ac.uk}
\affiliation{School of Math and Physics, University of Portsmouth, Portsmouth PO1 3QL, UK}
\affiliation{Institute of Cosmology \& Gravitation, University of Portsmouth, Portsmouth PO1 3FX, UK}
\affiliation{Institut f\"{u}r Quantenphysik and Center for Integrated Quantum Science and Technology (IQ\textsuperscript{ST}), Universität Ulm, D-89069 Ulm, Germany}
\author{Simon Laibacher}
\affiliation{Institut f\"{u}r Quantenphysik and Center for Integrated Quantum Science and Technology (IQ\textsuperscript{ST}), Universität Ulm, D-89069 Ulm, Germany}


\begin{abstract}
Multiphoton interference is an essential phenomenon at the very heart not only of fundamental quantum optics and applications in quantum information processing and sensing but also of demonstrations of quantum computational supremacy in boson sampling experiments relying only on linear optical interferometers.
	However, scalable boson sampling experiments with either photon-number states or squeezed states are challenged by the need to generate a large number of photons with fixed temporal and frequency spectra from one experimental run to another. Unfortunately, even the well established standard multiplexing techniques employed to generate photons with fixed spectral properties are affected by the detrimental effects of losses, spectral distorsions and reduction in purity.
	Here, we employ sampling correlation measurements in the photonic inner modes, time and frequency, at the interferometer input and output to ensure the occurrence of multiphoton interference even with pure states of input photons with random spectral overlap from one sample to another.  Indeed, by introducing a random multiplexing technique where photons are generated with random inner-mode parameters, it is possible to substantially enhance the probability to successfully generate samples and overcome the typical drawbacks in standard multiplexing.
	Remarkably, we demonstrate the classical hardness of the resulting problem of scattershot multiboson correlation sampling  based on this technique. 
	 Therefore, these results not only shed new light in the computational complexity of multiboson interference but also allow us to enhance the  experimental scalability of boson sampling schemes. Furthermore, this research provides a new exciting route toward future demonstrations of quantum computational supremacy with scalable experimental resources as well as future applications in quantum information processing and sensing beyond boson sampling.
\end{abstract}

\maketitle

Multiphoton interference is a fundamental phenomenon in quantum optics with numerous applications spanning from
quantum foundations \cite{Pan2012_Multiphotonentanglementinterferometry, Alley1986_newtypeEPR, Legero2004_QuantumBeatTwo, Tamma2015_MultibosonCorrelationInterferometry} to quantum information and sensing technologies   \cite{NielsenChuang, moreau2019imaging, slussarenko2019photonic, acin2017european, DAngelo2008_realmaximallypathentangled}.

Recently, the introduction of the boson sampling problem by Aaronson and Arkhipov (AABS)  \cite{Aaronson2011_computationalcomplexitylinear} and the following experimental realizations \cite{wangboson,Broome2013_Photonicbosonsampling,Crespi2013_Integratedmultimodeinterferometers,Tillmann2013_Experimentalbosonsampling,Spring2013_Bosonsamplingphotonic,Bentivegna2015_Experimentalscattershotboson,Wang2017_Highefficiencymultiphotonboson}  have provided the potential to achieve quantum computational supremacy without the need of a general-purpose quantum computer.
Indeed, AABS simply consists of the task of sampling from the probability distribution of all the possible ways of detecting $N$ interfering photons at the output of a linear interferometer (described by a unitary matrix $\mathcal{U}$  chosen randomly with respect to the Haar measure) after being injected in $N$ out of $M\sim N^2$ distinct fixed input ports \cite{Aaronson2011_computationalcomplexitylinear}.

A first enhancement in the scalability of boson sampling experiments was given by the introduction of Scattershot Boson Sampling (SBS) \cite{Aaronson2013_ScattershotBosonSamplingnew,Lund2014_BosonSamplingGaussian,Bentivegna2015_Experimentalscattershotboson}, where a fixed number $N$ of identical single photons are postselected in a random set of $N$ input ports by employing $M\sim N^2$ heralded spontaneous-parametric-downconversion (SPDC) sources. The recent introduction of the Gaussian boson sampling (GBS) problem, based on the use of non heralded sources of squeezed  states and therefore depending on matrix Hafnians instead of matrix permanents, has triggered interesting but still open questions on its computational hardness \cite{Hamilton2017GBS, Hamilton2019GBS, zhong2020quantum, Aaronson2020GaussianBosonSamplingBlog, renema2020simulability, qi2020regimes}. 

More recently, a variation of SBS with a random number of input photons and only a linear number of heralded SPDC sources was introduced  by additionally sampling at the interferometer input over all the possible input states of $N$ or more occupied input ports \cite{tamma2020boson}. This has allowed to achieve a probability of generating input samples which increases instead of decreasing with $N$, differently from SBS and GBS with a fixed number of postselected photons.

However, all these boson sampling schemes suffer from the experimental challenge to generate a large number of  input photons which are identical in their inner-mode properties in order to do not affect the computational hardness of the problem \cite{Tamma2015_MultibosonCorrelationInterferometry, Tichy2015, Shchesnovich2015,TammaLaibacherBSJMO}. Indeed, $N$ photons emitted by different  sources or by the same source at different times can differ in their frequency and temporal spectra, especially for $N > 50$ as required to achieve quantum computational supremacy \cite{Neville2017_Classicalbosonsampling}. This is indeed the case not only for single-photon emitters, such as diamond colour centers \cite{Babinec2010_diamondnanowiresinglephoton}, single molecules \cite{Lounis2000_Singlephotonsdemand} and quantum dots \cite{Shields2007_Semiconductorquantumlight,Michler2000_QuantumDotSinglePhoton} in boson sampling experiments with single photons but also for sources of squeezed light employed in GBS experiments \cite{shi2021gaussian,renema2020simulability,zhong2020quantum}.  

The experimental requirement for identical input photons in multiphoton interference experiments has been eased by introducing a sampling process called multiboson correlation sampling (MBCS) \cite{Laibacher2015_PhysicsComputationalComplexity}. Here, the input photons can differ in their inner mode parameters, e.g. frequencies or injection times, and sampling measurements resolving  the correspondent conjugate parameters are performed at the output \cite{Laibacher2015_PhysicsComputationalComplexity,Laibacher2018, Tamma2015_Multibosoncorrelationsampling,Tamma2014_Samplingbosonicqubits}.
Indeed, the advent of very fast single-photon detectors and high-precision single-photon spectrometers \cite{Legero2004_QuantumBeatTwo,GrimauPuigibert2017_HeraldedSinglePhotons,Tamma2015_MultibosonCorrelationInterferometry,Avenhaus2009_Fiberassistedsinglephotonspectrograph, Davis2017_Pulsedsinglephotonspectrometer,Polycarpou2012_AdaptiveDetectionArbitrarily,Gerrits2015_Spectralcorrelationmeasurements, Jin2015_SpectrallyresolvedHongOuMandel,Shcheslavskiy2016_Ultrafasttimemeasurements} have already made possible the realisation of MBCS experiments \cite{Wang2018ExpMBCS, Orre2019ExpMBCS}. 

Unfortunately,  even for MBCS schemes with non-identical input photons, a key experimental challenge is still the need  to generate a large number of input photons which states, although they may differ from each other, are not allowed to change between consecutive runs of the experiments.

One possible approach is the use of multiplexing to approximately generate SPDC states $\ket{\psi} =\otimes_{i=1}^{k}  [ \sqrt{1-\gamma^2} \sum_{n_i=0}^{\infty} \gamma^{n_i} \ket{n_i,n_i} ]$ of $k$ ``multiplexed'' modes with $n_i\geq 0$  photons and the same squeezing parameter $\gamma$\cite{Pittman2002_Singlephotonspseudodemand,Kaneda2015_Timemultiplexedheraldedsinglephoton,GrimauPuigibert2017_HeraldedSinglePhotons, hiemstra2020pure, Christ2012_Limitsdeterministiccreation}. Indeed, by increasing the number $k$ of inner modes in which each single photon can be generated by SPDC sources it is ideally possible to enhance the probability of single photon generation, while keeping the ratio between one photon pair events and two pair events constant. Only  $k=13$ inner modes with the optimal value $\gamma= \gamma_{\text{opt}} = 1/\sqrt{2}$ are already enough to reach, in principle, a probability $p=1-\bigl(1-\gamma_{\text{opt}}^2(1-\gamma_{\text{opt}}^2) \bigr)^k \cong 98 \% $ for single-photon generation with a substantial enhancement with respect to the  probability of $25 \%$ when no multiplexing is employed  \cite{Christ2012_Limitsdeterministiccreation}. 

Another promising approach based on demultiplexing of single photons emitted by a single quantum dot was also demonstrated experimentally \cite{Wang2017_Highefficiencymultiphotonboson}.

However, unfortunately, current multiplexing and demultiplexing techniques in the photonic inner modes, such as time and frequency \cite{Pittman2002_Singlephotonspseudodemand,Kaneda2015_Timemultiplexedheraldedsinglephoton,GrimauPuigibert2017_HeraldedSinglePhotons, hiemstra2020pure, Christ2012_Limitsdeterministiccreation, Wang2017_Highefficiencymultiphotonboson}, are threatened, especially for large photon numbers $N$, by the effect of losses, possible spectral distorsions and reduction in purity. This is due to the need of multiple optical elements such as switches, delay lines, storage cavities, or phase modulators which are employed to generate photons with fixed frequency and temporal properties, limiting also the range of exploitable multiplexed modes to scale up the probability in single-photon generation \cite{Christ2012_Limitsdeterministiccreation,GrimauPuigibert2017_HeraldedSinglePhotons, hiemstra2020pure}. Furthermore, for high values of squeezing parameters photon-number resolved detectors are required to ensure single-photon emissions.

Therefore, important questions remain to be addressed for future experiments at the intersection between quantum optics and quantum computational complexity.
Can classical hardness be achieved by employing single photons whose inner-mode properties change randomly from one sample to another ? Can the generation of input photons with random inner-mode properties  be turned from an experimental drawback to a resource for more scalable boson sampling realizations while only using a linear number of sources ? Can one take also advantage of multi-photon emission in multiplexing techniques without recurring to photon-number resolving detectors?


In this paper, we demonstrate the quantum computational complexity beyond any classical capabilities of quantum linear optics networks even with pure states of input photons which differ randomly in either their central frequencies or injection times from one experimental run to another. Remarkably, we also show how, in this case, a substantial scaling up of experimental implementations of boson sampling is achievable with the current technology. 

\begin{figure}[ht]
	\centering
	\includefig{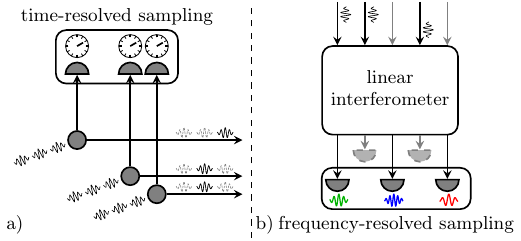}
	\vspace{-0.5cm}
	\caption{
		SMBCS implementation with input photons injected at random times.
	a) Random time multiplexing: Using time-resolved heralding of the idler photons emitted by an SPDC source pumped by a train of laser pulses, $N$ signal photons are generated in random but known time slots at each different sample.
		b) Multiphoton interferometer based on frequency-resolved correlation measurements: $N$ single photons, generated as in panel a, are injected into a passive linear network and detected using frequency-resolving detectors.
	}
	\label{fig:fig2}
\end{figure}

\begin{figure}[ht]
	\centering
	\includefig{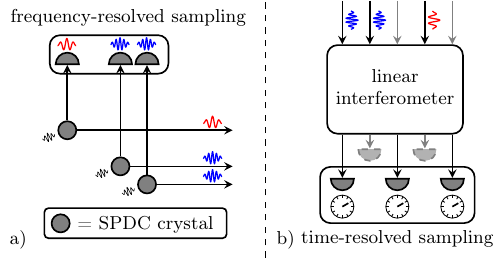}
	\caption{
		SMBCS implementation with input photons of random colors.
	a) Random frequency multiplexing: Using frequency-resolved heralding of the idler photons emitted by a pulsed SPDC source, $N$ signal photons are generated at random but known central frequencies at each different sample.
		b) Multiphoton interferometer based on time-resolved correlation measurements: $N$ single photons, generated as in panel a, are injected into a passive linear network and detected using time-resolving detectors.
	}
	\label{fig:fig1}
\end{figure}

In particular, we introduce a novel technique, \emph{random inner-mode multiplexing}, which employs heralding SPDC measurements to sample at the interferometer input over either the central times (Fig.~\ref{fig:fig2}a) 
or the central frequencies (Fig.~\ref{fig:fig1}a) of the input heralded photons.
Further, by sampling at the output in the respective conjugate parameters, instead of ``classically'' average over such information, it is possible to restore multiphoton indistinguishability at the detectors (see Fig.~\ref{fig:fig2}b and Fig.~\ref{fig:fig1}b, respectively).
The emergence of full multiphoton interference even with input photons with random spectral overlap is at the very heart of the classical hardness of this problem, named here \emph{scattershot multiboson correlation sampling (SMBCS)}, even in the approximate case.
Indeed, full quantum interference occurs at each experimental run for all possible injection times (frequencies) heralded at the input and all detected frequencies (times) at the output, even if these values vary randomly in consecutive measurements.  

Remarkably, by allowing the input photons to have either random central times or frequencies, losses, as well as possible spectral distortions and reduction in purity, due to additional optical elements needed to generate single photons in fixed inner modes are also avoided.
This allows to scale up SMBCS experiments to larger photon numbers in comparison to boson sampling, SBS, and MBCS experiments where the input photons are generated in fixed inner modes. Furthermore, no filtering for particular detection times or frequencies is necessary, since this technique takes advantage of all the possible measurement outcomes in the photonic inner modes.
%

Let us consider, in general, $N$ single photons injected into a set $\mathcal{S}$ of $N$ input ports of an arbitrary passive, linear optical network with a total of $M\sim N^2$ input and output ports.
Contrary to conventional boson sampling models, we allow these photons to have normalized spectra $\xi (\omega -\omega_s) \ee{\ii \omega t_s}$ with either different central frequencies $\omega_s$ or different central times $t_s$, with $s \in \mathcal{S}$, where $\xi (\omega -\omega_s)$ is a positive function with maximum at $\omega = \omega_s$ describing the photonic wave packets.
The overall input state can be written as the product state
\begin{equation}
	\ket{\mathcal{S}} \defeq \smashoperator[r]{\bigotimes_{s\in \mathcal{S}}}
	\ket{1;\omega_s,t_{s}}_{s}
	\smashoperator[r]{\bigotimes_{s \notin \mathcal{S}}}
	\ket{0}_{s}
	\label{eq:StateDefinition}
\end{equation}
of $M-N$  vacuum states in the unoccupied ports and $N$ single photon states
\begin{align}
	\ket{1;\omega_s,t_{s}}_{s} \defeq
	\int_{0}^{\infty} \d{\omega} \xi(\omega-\omega_s) \ee{+\ii \omega t_{s}} \hat{a}_{s}^{\dagger}(\omega) \ket{0}_{s}.
	\label{eq:SinglePhotonState}
\end{align}

\paragraph{Scattershot Multiboson Correlation Sampling (SMBCS) with input photons injected at random times.}
\label{sec:time_scattershot}

We consider first the case where the $N$ input photons in Eq.~\eqref{eq:SinglePhotonState} are generated with the same frequency ($\omega_s=\omega_0\ \forall s\in\mathcal{S}$) but at random different times $\{t_s\}$ from one sample to another by employing
the \textit{random time multiplexing (RTM)} technique in Fig.~\ref{fig:fig2}a.
Here, the input photons are the signal photons of pulsed SPDC sources heralded at random times by time-resolved detections of the idler photons.
Remarkably, by allowing random input times, this RTM technique is not affected by any losses associated with the optics required in boson sampling experiments based on standard time-multiplexing to generate the photons at fixed times (e.g.\ multiple delay lines and optical switchers) \cite{Pittman2002_Singlephotonspseudodemand,Kaneda2015_Timemultiplexedheraldedsinglephoton}, enhancing therefore substantially the experimental scalability to higher photon numbers.

Multiphoton interference at the interferometer output can be obtained by ``erasing'' any random time distinguishability via frequency-resolved measurements as depicted in Fig.~\ref{fig:fig2}b.
For photonic wave packets of a given bandwidth $\Delta \omega$, for example of Gaussian or rectangular shape,  this is ensured for small enough frequency resolution $\delta \omega$ according to the conditions
\begin{equation}
	\label{eq:condition_on_the_frequency_bin_size}
	\delta \omega  \ll \abs{t_{s}-t_{s'}}^{-1} \ \forall s,s'\in \mathcal{S} \quad \text{and} \quad \delta\omega \ll \Delta\omega.
\end{equation}
This includes the case where the input photons exhibit no overlap in their temporal spectra.

We can now define the ``Scattershot Multiboson Correlation Sampling'' (SMBCS) problem with input photons injected at random input times as  the task of generating an overall sample $[(\mathcal{D}, \{\omega_d\}_{d \in \mathcal{D}}),(\mathcal{S},\{t_{s}\}_{s \in \mathcal{S}})]$ determined by an input sample $(\mathcal{S},\{t_{s}\}_{s \in \mathcal{S}})$ of $N$ input ports $\mathcal{S}$ occupied with single photons with random injection times $\{t_s\}$  in addition to the output sample $(\mathcal{D}, \{\omega_d\}_{d \in \mathcal{D}})$  of $N$ detectors $\mathcal{D}$ detecting the $N$ photons at random frequencies $\{\omega_d\}$. 

Indeed, defining the annihilation operators $\hat{a}_d =\sum_{s\in\mathcal{S}}\mathcal{U}_{ds}\hat{a}_s$ at the detectors and  the permanent  $	\perm \mathcal{A} \defeq \sum_{\sigma\in\Sigma_N} \prod_{i=1}^{N} \mathcal{A}_{i\sigma(i)}$ of a matrix $\mathcal{A}$  as the sum over all possible permutations $\sigma$ from the symmetric group $\Sigma_N$ of order $N$, we find  the probability \footnote{Please see Appendix for a detailed derivation.}\cite{Laibacher2018}
\begin{equation}
	\label{eq:probabilityfrequency}
	\begin{split}
		P^{(\mathcal{D}, \mathcal{S})}_{\{\omega_{d}\},\{t_s\}} &= \delta\omega^{N}
		\matrixel[\big]{\mathcal{S}}{\smashoperator[r]{\prod_{d \in \mathcal{D}}} \hat{a}^{\dagger}_{d}(\omega_d)
		\smashoperator[r]{\prod_{d \in \mathcal{D}}} \hat{a}_d(\omega_d)}{\mathcal{S}} \\
		&\propto   \abs[\Big]{\perm \Big( \big[ \mathcal{U}_{ds} \ee{\ii \omega_d t_s} \big]_{\substack{d\in\mathcal{D}\\ s\in\mathcal{S}}} \Big) }^2,
	\end{split}
\end{equation}
of generating a sample $[(\mathcal{D},\{\omega_d\}_{d \in \mathcal{D}}),(\mathcal{S},\{t_{s}\}_{s \in \mathcal{S}})]$. The $N!$ permutations which are coherently added in the matrix permanents defining Eq.~\eqref{eq:probabilityfrequency} correspond to all possible multiphoton paths which bijectively connect the output ports $\mathcal{D}$ with the input ports $\mathcal{S}$.

In order to give evidence of the classical hardness of SMBCS, we recall  that,  for Haar random unitary matrices $U$ of size $M$, the elements $\mathcal{U}_{ds}$ of any submatrices of $U$ of size $N \ll M$ are independent and identically distributed (i.i.d.) Gaussian random variables with a standard normal
distribution $\mathcal{N}(0; 1)_{\mathcal{C}}$ with mean $0$ and variance $1$ \cite {Aaronson2011_computationalcomplexitylinear}. Therefore, the same is valid for the entries $\mathcal{U}_{ds} \ee{\ii \omega_d t_s}$ of the $N\times N$ submatrices  in Eq. (\ref{eq:probabilityfrequency}),  independently of the values of $t_s$ and $\omega_d$, since the additional phase factors $\ee{\ii \omega_d t_s}$
 only rotate the elements $\mathcal{U}_{ds}$ in the complex plane and the distribution
$\mathcal{N}(0; 1)_{\mathcal{C}}$ is symmetric around the origin \cite{Laibacher2015_PhysicsComputationalComplexity}. Permanents of matrices with i.i.d. Gaussian random entries as in Eq. (\ref{eq:probabilityfrequency}) are strongly believed to be classically hard to estimate \cite{Aaronson2011_computationalcomplexitylinear}, implying that approximate SMBCS with random input injection times is at least as classically hard as AABS and MBCS \cite{Laibacher2015_PhysicsComputationalComplexity}. Given $k$ possible ``multiplexed'' injected times, the SMBCS problem has additionally the ``bonus'' that the number of possible samples  is $k^N$ times larger in comparison to MBCS experiments with photons injected at fixed times.

Furthermore, the dramatic reduction of losses in the RTM technique enables to scale up boson sampling experiments by avoiding the need to generate photons at fixed times via standard time multiplexing \cite{Pittman2002_Singlephotonspseudodemand,Kaneda2015_Timemultiplexedheraldedsinglephoton}.
In addition, by increasing the pulsed pump laser repetition rate $f_{\text{p}}$ or by decreasing the frequency resolution $\delta\omega$ of the detectors it is possible to substantially increase the maximum number of pulses $k \ll f_{\text{p}}/\delta\omega$, therefore enhancing the single-photon generation probability.
One may also extend this technique to the use of a cw pump laser which would lead to a maximum number of RTM time bins $k \ll (\delta t \delta\omega)^{-1}$ determined by the time resolution $\delta t$ and the frequency resolution $\delta\omega$ of the detectors.

\paragraph{Scattershot Multi-Boson Correlation Sampling (SMBCS) with input photons of random colors.}
\label{sec:frequency_scattershot}

We consider now the case where the $N$ input photons in Eq.~\eqref{eq:SinglePhotonState} are injected into the interferometer at the same time $t_s=t$ but can have random colors (generally, $\omega_s \neq \omega_{s'}$ if $s \neq s'$), by using the technique of \emph{random frequency multiplexing (RFM)} in Fig.~\ref{fig:fig1}a.
Here, SPDC sources are used to generate single photons at random, but known, central frequencies through frequency-resolved heralding with a single-photon spectrometer \cite{GrimauPuigibert2017_HeraldedSinglePhotons}.
This technique differs from the standard frequency multiplexing technique introduced in Ref.~\cite{GrimauPuigibert2017_HeraldedSinglePhotons}, where a phase modulator is used to shift the random frequencies of the heralded single photons to a fixed value.
Indeed, the RFM technique allows us to sample from all the possible random frequencies of the input photons, which we can assume for simplicity to occur equally likely.
Additionally, at the interferometer output in Fig.~\ref{fig:fig1}b a sampling process occurs both in the output ports and the detections times  the photons are detected at with detector integration times 
\begin{equation}
	\label{eq:conditionsintegrationtime}
\delta t \ll \abs{\omega_s - \omega_{s'}}^{-1}  \ \forall s, s' \in \mathcal{S}\text{\quad and\quad} \delta t \ll 1/\Delta \omega
\end{equation}
to ensure multiphoton indistinguishability. Indeed, this condition corresponds to Eq.~\eqref{eq:condition_on_the_frequency_bin_size} by interchanging the conjugate variables frequency and time.

We can now define SMBCS with input photons of random colors as the task of generating an overall sample $[(\mathcal{D}, \{t_d\}_{d \in \mathcal{D}}),(\mathcal{S},\{\omega_{s}\}_{s \in \mathcal{S}})]$ determined by input sample $(\mathcal{S},\{\omega_{s}\}_{s \in \mathcal{S}})$ of $N$ input ports $\mathcal{S}$ occupied with single  photons of random central frequencies $\{\omega_s\}$  in addition to the output sample $(\mathcal{D}, \{t_d\}_{d \in \mathcal{D}})$  of $N$ detectors $\mathcal{D}$ detecting the $N$ photons at random detection times $\{t_d\}$.


Indeed, by interchanging now times and
frequencies in Eq. (\ref{eq:probabilityfrequency}), the probability 
\begin{equation}
	\label{eq:probabilityrectangle}
	P^{(\mathcal{D},\mathcal{S})}_{\{t_d\},\{\omega_s\}}\propto \abs[\Big]{\perm \Big( \big[ \mathcal{U}_{ds} \ee{\ii \omega_s t_d} \big]_{\substack{d\in\mathcal{D}\\ s\in\mathcal{S}}} \Big) }^2
\end{equation}
of generating the sample $[(\mathcal{D}, \{t_d\}_{d \in \mathcal{D}}),(\mathcal{S},\{\omega_{s}\}_{s \in \mathcal{S}})]$ is defined again by permanents of matrices with i.i.d.\ Gaussian random entries \cite{Laibacher2015_PhysicsComputationalComplexity}. Analogously to the previous case, this implies that approximating SMBCS with input photons of random colors is at least as classically hard as AABS and MBCS \citep{Laibacher2015_PhysicsComputationalComplexity}.

%

Furthermore, the RFM technique allows to avoid the losses connected with the need in standard frequency multiplexing of phase modulators to shift the photons heralded at a random frequency to a fixed frequency, enabling a scaling up of boson sampling experiments to higher photon numbers \cite{GrimauPuigibert2017_HeraldedSinglePhotons}.

In addition, while in standard frequency multiplexing the maximum number $k_{max}$ of central frequencies is limited by the maximum frequency shift allowed by the phase modulators, in RFM $k_{max}$ is only limited by the detector resolution according to Eq.~\eqref{eq:conditionsintegrationtime}. As an example, given a frequency resolution of $\SI{1}{\giga\hertz}$ and a time resolution of $\SI{10}{\pico\second}$, it is possible to sample already over $k \sim 10$ different central frequencies.

\paragraph{SMBCS with more than one single photon per input channel}.
So far we have considered SMBCS with one single photon in only one of the $k$ multiplexed modes at each of the $N$ input channels. We exploit now the case where $n_s \geq 1$ single photons, each in a different multiplexed mode, are allowed to be injected at each input channel $s \in \mathcal{S}$, leading to an input spatial sample $\mathcal{S'}$ of $N'\geq N$ occupied input channels where each  channel $s$ is counted $n_s$ times.  In such a case, one can define the SMBCS samples $[(\mathcal{D}', \{t_{d'}\}_{d' \in \mathcal{D'}}),(\mathcal{S'},\{\omega_{s'}\}_{s' \in \mathcal{S'}})]$, for input photons with random colours, and  $[(\mathcal{D}',\{\omega_{d'}\}_{d' \in \mathcal{D}'}),(\mathcal{S'},\{t_{s'}\}_{s' \in \mathcal{S'}})]$, for input photons at random times, which are  detected in a random set $\mathcal{D}'$ of $N'$ distinct output channels by resolving the corresponding conjugate parameters. The corresponding detection probabilities $P^{(\mathcal{D'},\mathcal{S'})}_{\{t_{d'}\},\{\omega_{s'}\}}$ and $P^{(\mathcal{D'}, \mathcal{S'})}_{\{\omega_{d'}\},\{t_{s'}\}}$ for such samples in Eqs.~\eqref{eq:probabilityrectangle} and \eqref{eq:probabilityfrequency}, respectively, depend now on permanents of $N' \times N'$ matrices with $n_s$ columns for each input channel $s\in \mathcal{S}$, containing the same single-photon interferometer transition amplitudes $\mathcal{U}_{d's}$ but different phase factors associated to the different $n_s$ input injection times or frequencies.
Since the $N$ sources produce at least one photons, such matrices contain at least $N\times N$ entries which are i.i.d.\ Gaussian random variables, which is again a strong evidence that such a problem is at least as classical hard as AABS and MBCS \cite{Aaronson2011_computationalcomplexitylinear, Laibacher2015_PhysicsComputationalComplexity}. Remarkably, by considering  additional samples where more than one single photon can be injected per channel the number of possible samples further increases exponentially with the number of multiplexed modes.

Interestingly, the total number  $N'$ of photons is always proportional to the number $N$ of independent sources leading to negligible photon bunching events at the interferometer $M \sim N^2$ output ports for large values of $N$, avoiding therefore the need of photon-number resolving detectors \cite{Aaronson2011_computationalcomplexitylinear}.  Remarkably, this also allows us to distinguish events where one of the multiplexed input modes is occupied with more than one photon without again employing photon-number resolving detectors as in standard multiplexing techniques. 

Furthermore, the possibility of taking into advantage sampling events with more than $N$ single photons has the potential to enhnance the robustness of SMBCS to losses.

\paragraph{Discussion.}

We have introduced and shown the classical hardness even in the approximate case of the so-called  scattershot multiboson correlation sampling (SMBCS) problem employing a linear interferometer with input photons which, differently from previous schemes, do not need to be generated with fixed frequencies or injection times from one experimental run to another.

 Indeed, SMBCS relies on sampling  over either the random central frequencies or injection times of the input photons generated by using the newly introduced random inner-mode multiplexing (RIMM) technique (Figs.~\ref{fig:fig1} and \ref{fig:fig2}). Remarkably, this technique also overcomes two main drawbacks in standard multiplexing techniques imposed by the need of  tuning the SPDC generated photons to fixed inner-mode parameters from one experimental run to another:  losses associated with the necessary auxiliary optics;  limitation in the range of exploitable multiplexed frequencies or time of emissions. 

Therefore, SMBCS allows both a scaling up in experimental realizations and an exponential increase of the number of samples with the number of multiplexed modes with respect to current schemes where the input single photons are required to have fixed inner modes at each experimental run. 

Interestingly, the RIMM technique can be applied in general to schemes with input photons with any random inner mode parameters in addition to the central frequencies and injection times considered here, e.g. transverse momentum or orbital angular momentum. Indeed, by resolving at the interferometer output the correspondent photonic conjugate parameters,  multi-photon indistinguishability is retrieved independently of the input inner modes from one experimental run to another.

One can  further take advantage of an additional number of samples, again exponential with the number of multiplexed modes, by allowing more than one mode per channel to be occupied with single photons and without requiring photon-number resolving detectors. This motivates an interesting question: would it be possible to demonstrate SMBCS quantum computational supremacy by keeping the number of sources fixed independently of $N$ or even using a single source with $N$ injected single photons in $N$ different multiplexed modes in a single input port?  Further studies of the computational complexity of SMBCS interferometers in this or related scenarios will be surely the focus of future works. 

Finally, these results not only can be exploited to enhance the scalability of any boson sampling scheme with either photon number states or squeezed states with a fixed or random number of detected photons  \cite{tamma2020boson, Hamilton2017GBS, Lund2014_BosonSamplingGaussian}, but can also inspire new platforms for applications in quantum sensing and information processing beyond boson sampling, where experimental losses are minimized and the quantum information stored in the spectra of experimentally non-identical photons is exploited in its fullness \cite{Tamma2015_MultibosonCorrelationInterferometry, Zimmermann2017_Whichroledoes}.

\begin{acknowledgments}
The authors are grateful to S. Aaronson, A. Arkhipov, C. Dewdney, M. Foss-Feig, K. Jacobs, S. Kolthammer, A. Laing, F. Sciarrino, and P. Walther for discussions related to this work.
V. T. is also thankful to W. P. Schleich for the time passed at the Institute of Quantum Physics in Ulm where some of the ideas behind this work started to flourish.

Research was partially sponsored by the Army Research Laboratory and was accomplished under Cooperative Agreement Number W911NF-17-2-0179.
The views and conclusions contained in this document are those of the authors and should not be interpreted as representing the social policies, either expressed or implied, of the Army Research Laboratory or the U.S. Government.
The U.S. Government is authorized to reproduce and distribute reprints for Government purposes notwithstanding any
copyright notation herein.
S.L. acknowledges support by a grant from the Ministry of Science, Research and the Arts of Baden-W\"urttemberg (Az: 33-7533-30-10/19/2).
\end{acknowledgments}

\begin{widetext}
\textbf{ Appendix: Frequency correlations of \texorpdfstring{$N$}{N} photons}
\setcounter{equation}{0}
\setcounter{figure}{0}
\setcounter{table}{0}
\makeatletter
\renewcommand{\theequation}{S\arabic{equation}}
\renewcommand{\thefigure}{S\arabic{figure}}

The detection of a photon in output port $d \in \mathcal{D}$ and at frequency $\omega_d$ can be described by the application of the annihilation operator $\hat{a}_d(\omega_d)$ to the output state of the interferometer.
Therefore, the probability density to measure the outcome $(\mathcal{D},\{\omega_d\})$ is given by the correlation function
\begin{equation}
	\label{eq:supp:correlationfunction}
	\matrixel[\big]{\mathcal{S}}{\smashoperator[r]{\prod_{d \in \mathcal{D}}} \hat{a}^{\dagger}_{d}(\omega_d)
			\smashoperator[r]{\prod_{d \in \mathcal{D}}} \hat{a}_d(\omega_d)}{\mathcal{S}}
\end{equation}
evaluated for the input state (Eqs.~$(1)$ and $(2)$ in the main paper)
\begin{equation}
	\label{eq:supp:StateDefinition}
	\ket{\mathcal{S}} \defeq \smashoperator[r]{\bigotimes_{s\in \mathcal{S}}}
	\ket{1;\omega_s,t_{s}}_{s}
	\smashoperator[r]{\bigotimes_{s \notin \mathcal{S}}}
	\ket{0}_{s}
\end{equation}
with
\begin{equation}
	\label{eq:supp:SinglePhotonState}
	\ket{1;\omega_s,t_{s}}_{s} \defeq
	\int_{0}^{\infty} \d{\omega} \xi(\omega-\omega_s) \ee{+\ii \omega t_{s}} \hat{a}_{s}^{\dagger}(\omega) \ket{0}_{s}.
\end{equation}
Naturally, the structure of this correlation function strongly depends on the passive, linear optical network.
Its effect can be described as a linear transformation which connects the output mode operators $\hat{a}_{d}(\omega)$ with the input mode operators $\hat{a}_{s}(\omega)$ via the single-photon transition amplitudes $\mathcal{U}_{ds}$.
Since the unoccupied input ports $s\not\in \mathcal{S}$ do not contribute to the correlations, we can for given $\mathcal{S}$ and $\mathcal{D}$ effectively write
\begin{equation}
	\label{eq:supp:lineartransformation}
	\hat{a}_d(\omega_d) = \sum_{s\in \mathcal{S}}\mathcal{U}_{ds} \hat{a}_s(\omega_d).
\end{equation}
The probability to detect the $N$ photons in the input state in Eq.~(\ref{eq:supp:StateDefinition}) at the output ports $\mathcal{D}$ and in the detected frequency intervals $I(\{\omega_d\})\defeq \bigotimes_{d\in\mathcal{D}}[\omega_d -\delta\omega/2, \omega_d +\delta\omega/2]$ consequently is
\begin{equation}
	\label{eq:supp:probability_frequency}
			P^{(\mathcal{D}, \mathcal{S})}_{\{\omega_{d}\},\{t_s\}} =
			\int_{I(\{\omega_d\})}\prod_{d\in\mathcal{D}}\d{\omega_d}\matrixel[\big]{\mathcal{S}}{\smashoperator[r]{\prod_{d \in \mathcal{D}}} \hat{a}^{\dagger}_{d}(\omega_d)
			\smashoperator[r]{\prod_{d \in \mathcal{D}}} \hat{a}_d(\omega_d)}{\mathcal{S}}.
		\end{equation}
Using the notation $\|\ket{\psi}\|  \defeq \braket{\psi}{\psi}$ and Eq.~\eqref{eq:supp:lineartransformation}, we can rewrite the correlation function, Eq.~\eqref{eq:supp:correlationfunction}, as
\begin{equation}
			\matrixel[\big]{\mathcal{S}}{\smashoperator[r]{\prod_{d \in \mathcal{D}}} \hat{a}^{\dagger}_{d}(\omega_d)
			\smashoperator[r]{\prod_{d \in \mathcal{D}}} \hat{a}_d(\omega_d)}{\mathcal{S}}
		     = \left\| \prod_{d \in \mathcal{D}} \hat{a}_d(\omega_d) \ket{\mathcal{S}} \right\|^2
		     = \left\| \prod_{d\in \mathcal{D}} \sum_{s_d \in \mathcal{S}} \mathcal{U}_{ds_d} \hat{a}_{s_d}(\omega_d) \ket{\mathcal{S}} \right\|^2
		     = \Biggl\| \sum_{\{s_d\}\in \mathcal{S}^{N}} \prod_{d \in \mathcal{D}} \mathcal{U}_{d s_d} \hat{a}_{s_d}(\omega_d) \ket{\mathcal{S}} \Biggr\|^2.
\end{equation}
This expression can be further simplified by noting that due to the structure of the state $\ket{\mathcal{S}}$, Eq.~\eqref{eq:supp:StateDefinition}, only those terms contribute, in which each of the $N$ annihilation operators $\hat{a}_{s}(\omega)$, $s\in\mathcal{S}$, appears exactly once.
Denoting the set of all permutations of $N$ elements, the symmetric group of order $N$, as $\Sigma_N$ and recalling Eq.~\eqref{eq:supp:StateDefinition}, the correlation function becomes
\begin{equation}
			\matrixel[\big]{\mathcal{S}}{\smashoperator[r]{\prod_{d \in \mathcal{D}}} \hat{a}^{\dagger}_{d}(\omega_d)
			\smashoperator[r]{\prod_{d \in \mathcal{D}}} \hat{a}_d(\omega_d)}{\mathcal{S}}
		     = \left\| \sum_{\sigma \in \Sigma_N} \prod_{d \in \mathcal{D}} \mathcal{U}_{d \sigma(d)} \hat{a}_{\sigma(d)}(\omega_d) \ket{\mathcal{S}} \right\|^2
		     = \left\| \sum_{\sigma \in \Sigma_N} \prod_{d \in \mathcal{D}} \mathcal{U}_{d \sigma(d)} \hat{a}_{\sigma(d)}(\omega_d) \ket{1;\omega_{\sigma(d)},t_{\sigma(d)}]}_{\sigma(d)} \right\|^2.
\end{equation}
With the help of the definition of the single-photon states $\ket{1;\omega_s,t_s}$ in Eq.~\eqref{eq:supp:SinglePhotonState}, and recalling the definition $perm \mathcal{A} = \sum_{\sigma\in \Sigma_N} \prod_{i=1}^{N} \mathcal{A}_{i\sigma(i)}$ OF THE permanent of an $N\times N$ matrix with elements $\mathcal{A}_{ij}$, this can be simplified to
\begin{equation}
	\begin{split}
			\matrixel[\big]{\mathcal{S}}{\smashoperator[r]{\prod_{d \in \mathcal{D}}} \hat{a}^{\dagger}_{d}(\omega_d)
			\smashoperator[r]{\prod_{d \in \mathcal{D}}} \hat{a}_d(\omega_d)}{\mathcal{S}}
			&= \left\| \sum_{\sigma \in \Sigma_N} \prod_{d \in \mathcal{D}} \mathcal{U}_{d \sigma(d)} \xi(\omega_d-\omega_{\sigma(d)})  \ee{\ii \omega_d t_{\sigma(d)}}\ket{0} \right\|^2
			= \left| \sum_{\sigma \in \Sigma_N} \prod_{d \in \mathcal{D}} \mathcal{U}_{d \sigma(d)} \xi(\omega_d-\omega_{\sigma(d)}) \ee{\ii \omega_d t_{\sigma(d)}}\right|^2 \\
		     &= \abs[\Big]{\perm \Big( \big[ \mathcal{U}_{ds} \xi(\omega_d-\omega_s) \ee{\ii \omega_d t_s} \big]_{\substack{d\in\mathcal{D}\\ s\in\mathcal{S}}} \Big) }^2,
	\end{split}
	\label {eq:supp:permanent_definition},
\end{equation}
which, by considering photons with frequency spectra of width $\Delta\omega$ and identical central frequencies $\omega_s=\omega_0 \ \forall s\in\mathcal{S}$, reduces to 
\begin{equation}
	\begin{split}
			\matrixel[\big]{\mathcal{S}}{\smashoperator[r]{\prod_{d \in \mathcal{D}}} \hat{a}^{\dagger}_{d}(\omega_d)
			\smashoperator[r]{\prod_{d \in \mathcal{D}}} \hat{a}_d(\omega_d)}{\mathcal{S}}= \Big[ \prod_{d \in \mathcal{D}}  \xi(\omega_d-\omega_0) \Big]\;
	\abs[\Big]{\perm \Big( \big[ \mathcal{U}_{ds} \ee{\ii \omega_d t_s} \big]_{\substack{d\in\mathcal{D}\\ s\in\mathcal{S}}} \Big) }^2.
		\end{split}
	\label {eq:supp:permanent_definition final}
\end{equation}

Then, under the conditions
\begin{equation}
	\label{eq:supp:condition_on_the_frequency_bin_size}
	\delta \omega \abs{t_{s}-t_{s'}} \ll 1 \ \forall s,s'\in \mathcal{S} \quad \text{and} \quad \delta\omega \ll \Delta\omega
\end{equation}
from Eq.~($5$) in the main paper, which ensure that the detectors do not “classically average” over the correlations, the probability in Eq.~\eqref{eq:supp:probability_frequency} takes the form
\begin{equation}
	\label{eq:supp:probability_frequency_final}
	P^{(\mathcal{D}, \mathcal{S})}_{\{\omega_{d}\},\{t_s\}} = \
	\delta\omega^{N} \Big[ \prod_{d \in \mathcal{D}}  \xi(\omega_d-\omega_0) \Big]\;\abs[\Big]{\perm \Big( \big[ \mathcal{U}_{ds} \ee{\ii \omega_d t_s} \big]_{\substack{d\in\mathcal{D}\\ s\in\mathcal{S}}} \Big) }^2,
\end{equation}
where $\Big[ \prod_{d \in \mathcal{D}}  \xi(\omega_d-\omega_0) \Big] = \Delta\omega^{-N}$ if we consider for example input photons of rectangular shape.


\end{widetext}
\bibliography{SMBCS_tamma}{}
\bibliographystyle{ieeetr}
\bibliographystyle{apsrev4-1}
\end{document}